\documentclass[aps,pra,twocolumn]{revtex4}
\usepackage{amsmath}
\usepackage{amssymb}

\begin{document}
\title{Remark on laser linewidth hazard in opto-mechanical cooling}    
\author{Lajos Di\'osi}
\email{diosi@rmki.kfki.hu}
\homepage{www.rmki.kfki.hu/~diosi} 
\affiliation{
Research Institute for Particle and Nuclear Physics, H-1525 Budapest 114, POB 49, Hungary}
\affiliation{
Department of Physics, Technion --- Israel Institute of Technology, 32000 Haifa, Israel} 

\date{\today}

\begin{abstract}
I discuss the robustness of the pumped cavity dynamics against 
phase diffusion of the laser and conclude that opto-mechanical
cooling has extreme sensitivity compared to laser cooling of atoms. 
Certain proposals of ground state opto-mechanical cooling by single cavity
would require an unrealistic sharp laser linewidth
or equivalently, a very low level of phase noise. A systematic way to
cancel classical excess phase noise is the interferometric twin-cavity pumping, initiated 
for optically trapped macro-mirrors of future gravitational-wave detectors.
\end{abstract}

\maketitle
Very recently, numerous works \cite{Wil07,Mar07,Dan08,Gro08,Gen08,Bha08,Gen08x} 
have predicted or suggested that laser cooling can bring a nano-mechanical oscillator (nano-mirror) 
close to its quantum ground state. It is hard to miss the conceptual similarities between opto-mechanical
cooling and the standard (e.g. Doppler) laser cooling of atoms. This time the cooled object is the 
spatial motion of the mirror instead of the atom, and the refrigerator is an optical cavity oscillator 
instead of the atom's internal two-level system. The same way in both cases, the refrigerator has high 
optical excitation frequencies, the object has thermally excited modes of low (radio) frequencies; therefore the 
object-refrigerator coupling is practically missing. What really turns
the atomic two-level system or the cavity into a refrigerator is the external laser field.
Typical limitations of atom cooling are determined by the spontaneous decay rate $\kappa$ of the atom,
hence laser imperfectness (linewidth $\Gamma_l$, basically) does not influence the mechanism as far as:
\begin{equation}
\Gamma_l \ll \kappa~.
\end{equation}
I will conjecture that for opto-mechanical cooling the condition becomes:
\begin{equation}
n\Gamma_l \ll \kappa~,
\end{equation}
where $\kappa$ is the decay rate of the cavity field. This condition 
puts a fatally stronger limit on $\Gamma_l$ because of the large factor $n$:  
the steady state excitation number of the pumped cavity. Violating this condition will not
invalidate opto-mechnical cooling in general. Ground-state cooling, however, becomes more problematic
than thought before.

Let us follow the standard theory and Langevin equation formalism, shared by most of the cited works, 
to introduce the time-dependent phase $\phi$ of the laser field into the equation of the 
cavity mode absorption operator: 
\begin{equation}
\dot a =-(\kappa+i\Delta)a + Ee^{-i\phi} +\sqrt{2\kappa}a^{in}+\dots
\end{equation}
where $\Delta>0$ is the detuning of the cavity mode, $E$ is proportional to the pump field
\cite{Wil07,Gen08,Gen08x,VitPat07}.
The third term on the r.h.s. denotes the quantum noise coming from the vacuum environment 
at $T=0$ \cite{foot}: 
\begin{equation}
\langle a_{in}(t)a_{in}^\dagger(s)\rangle=\delta(t-s)~,
\end{equation}
and the ellipses stand for the coupling to the position of the mirror. The diffusion of the phase $\phi$ 
of the laser light is determined by the white-noise correlation:
\begin{equation}
\langle\dot\phi(t)\dot\phi(s)\rangle=2\Gamma_l\delta(t-s)~.
\end{equation}
This standard ansatz corresponds to a flat power spectrum $S(\omega)=2\Gamma_l$ of frequency fluctuations.
The assumption will be refined later.
We perform two subsequent canonical transformations $a\rightarrow a e^{-i\phi}$ and
$a\rightarrow a+\alpha$ where \hbox{$\alpha=E/(\kappa+\Gamma_l+i\Delta)$} is the large mean amplitude 
and $a$ becomes a small perturbation around it. We obtain:
\begin{equation}
\dot a =-(\kappa+\Gamma_l+i\Delta)a +i\alpha\dot\phi +\sqrt{2\kappa}a^{in}+\dots
\end{equation}
Note that we have approximated the term $i(\alpha+a)\dot\phi$ by $i\alpha\dot\phi$.     

With the choice of small detuning $\Delta>0$, our refrigerator becomes equivalent with a 
central oscillator of low frequency $\Delta$, that can have strong, even
resonance, coupling to the mirror's mechanical oscillation. 
One would think that we got the low-frequency refrigerator operating at $T=0$ almost for free. 
In reality, however, the main resource of cooling is the perfect periodic driving field. 
The relevant imperfectness is the finite linewidth $\Gamma_l$ of the laser. Indeed,
we must assure in eq.~(6) that the contribution of the phase noise (5) remain much less than 
the contribution of the quantum noise (4), which means $\vert\alpha\vert^2\Gamma_l\ll\kappa$.
This is just our condition (2), since $\vert\alpha\vert^2=n$ for large $\alpha$. 
If the condition is
not satisfied, the phase noise will impose an effective non-zero temperature on the cavity oscillator
and it can not act as a refrigerator to $T=0$ anymore. Let us ignore the structural difference
between the noises $a^{in}$ and $\dot\phi$, and imagine that the contribution of the large phase noise (5) 
is equivalent with the contribution of the quantum noise at a certain high (effective) temperature $T$:
\begin{equation}
\langle a_{in}(t)a_{in}^\dagger(s)\rangle=\frac{k_B T}{\hbar\Delta}\delta(t-s)~.
\end{equation}
Then the following estimation can be made for the temperature of the effective cavity mode, 
caused by the phase noise:
\begin{equation}
k_B T\sim \hbar\Delta\frac{n\Gamma_l}{\kappa}~.
\end{equation}

Finally, let us consider the concrete magnitudes of the parameters considered, e.g., in ref.~\cite{Gen08}.
Accordingly, we take $\kappa\sim\Delta\sim10$MHz, the $50$mW laser power at $1064$nm wavelength yields 
$E\sim10^{13}$Hz, and we are led to $n=\vert\alpha\vert^2\sim10^{10}-10^{11}$. This huge number would, 
via condition (2), impose a request of $\Gamma_l$ less than $10^{-4}-10^{-3}$Hz! 
This range is far from being available now. The recent experimental work \cite{Schetal07} estimates 
the deteriorating influence of phase noise in the alternative regime $\kappa\ll\Delta$ and for a stiffer
oscillator. The lowest achievable excitation scales with $\sqrt{T\Gamma_l}$. Ground-state cooling of a 
$40\mathrm{MHz}$ oscillator from a cryogenic temperature $T$ will still require $10^{5}$ times smaller 
noise intensity than the value ($\sim400\mathrm{kHz}$) observed in the experiment.

As anticipated above, we refine the standard ansatz (5). Since the detuning $\Delta$ is used in resonance with
the high quality oscillator, it is only the frequency noise spectrum $S(\omega)$ in a narrow band around 
$\Delta$ that matters \cite{Schetal07}. In reality, the strength $S(\Delta)$ can be, or can be made, much different from 
the linewidth $\Gamma_l$. Our calculations and considerations can invariably be retained just we replace 
$\Gamma_l$ by $S(\Delta)$. Obviously, the formulated demands should concern $S(\Delta)$ and its vicinity 
rather than the whole spectrum $S(\omega)$, rather than the linewidth $\Gamma_l$. The reduction of phase 
noise in a narrow band above 1MHz might be a less difficult task than the reduction of the total spectrum
and linewidth.  

I have restricted my calculations and arguments for the behaviour of the cavity oscillator (refrigerator). 
In mind, I had the back-action (self-cooling) method, while the active feed-back control (cold damping) 
method may turn out less vulnerable by the laser instabilities. Clearly, the coupled linearized 
quantum Langevin equations must be extended and solved exactly for the steady state in the presence of 
the phase noise term. It is likely that the full `cost' of the ground-state opto-mechanical refrigerator 
will contain the cost of extreme laser stability. 

Nonetheless, an idea that emerged in gravitational-wave interferometry might neutralize the laser
instability for nano-mirror cooling as well. Consider two identical cavities pumped by the 
same laser at the same phase. Then we have two cavity amplitudes $a$ and $b$ of identical
behaviour, including the identity $\alpha=\beta$ of their respective steady state mean amplitudes. 
By introducing the modes $(a-b)/\sqrt{2}\rightarrow a$ and $(a+b)/\sqrt{2}\rightarrow b$, the 
``differential'' mode satisfies:
\begin{equation}
\dot a =-(\kappa+\Gamma_l+i\Delta)a +i\dot\phi a +\sqrt{2\kappa}a^{in}+\dots
\end{equation}
Note that the large noise term $i\alpha\dot\phi$ has cancelled, we have to retain the small one $i\dot\phi a$.
This mode is a $T=0$ refrigerator, indeed! Its performance is only limited by the constraint (1),
instead of (2).
The coupling of the mirror motion to this mode is straightforward if, e.g., we
use a shared movable end mirror, silvered on both sides, between the two cavities. Such setups have been
suggested and analysed for gravitational-wave interferometer macro-mirrors to cancel the influence of 
laser instabilities \cite{Cor05} and to project quantum mechanical tests \cite{Cor06,Wip08}.
The double-cavity concept itself exists for nano-mirrors as well, 
so far unrelated to the laser noise issue \cite{Bha08}, and with independent pumpings \cite{VitPat07}.
To implement interferometric twin-cavities in ground state cooling of nano-mirrors seems a 
reasonable, if not unavoidable, next step.

I thank Marcus Aspelmeyer, David Vitali and Chris Wipf for useful correspondence.
This work was supported in part at the Technion by a fellowship from the Lady Davis Foundation,
and by the Hungarian OTKA Grant No. 49384.

\vfill

\begin{thebibliography}{99}
\bibitem{foot} The temperature of the environment can always be ingnored at optical frequencies.
\end{thebibliography}
\end{document}